\documentclass{ws-ijmpd}
\usepackage[super,compress]{cite}

\usepackage{graphicx}
\usepackage{mathtools, amssymb}
\usepackage[title]{appendix}
\usepackage{nomencl}
\usepackage[colorlinks=true, linkcolor=black, urlcolor=blue, citecolor=blue]{hyperref}
\makenomenclature

\newcommand{\overbar}[1]{\mkern 1.5mu\overline{\mkern-1.5mu#1\mkern-1.5mu}\mkern 1.5mu}
\newcommand{\eref}[1]{\eqref{#1}} 
\newcommand{\Eref}[1]{Eq.~(\ref{#1})} 
\newcommand{\Sref}[1]{Section~\ref{#1}} 
\newcommand{\Fref}[1]{Fig.~\ref{#1}} 
\newcommand{\Tref}[1]{Table~\ref{#1}} 
\newcommand{\Cref}[1]{Ref.~\refcite{#1}} 
\newcommand{\firsttabline}{\hline\noalign{\smallskip}}
\newcommand{\bodytabline}{\noalign{\smallskip}}
\newcommand{\lastabline}{\noalign{\smallskip}\hline}

\newcommand{\LM}{\mathcal{L}_\text{M}}
\newcommand{\LF}{\mathcal{L}_\text{F}}
\newcommand{\LEH}{\mathcal{L}_\text{EH}}
\newcommand{\UM}{U_\text{M}}
\newcommand{\UB}{U_\text{B}}
\newcommand{\eM}{\varepsilon^\text{M}}
\newcommand{\eB}{\varepsilon^\text{B}}
\newcommand{\sM}{\sigma^\text{M}}
\newcommand{\sB}{\sigma^\text{B}}
\newcommand{\SF}{\mathcal{S}_\text{F}}

\newcommand{\Rthree}{R^\text{3D}}

\newcommand{\ethree}{\varepsilon^\text{3D}}
\newcommand{\e}{\varepsilon}
\newcommand{\SEH}{\mathcal{S}_\text{EH}}

\newcommand{\calO}{\mathcal{O}}
\newcommand{\calU}{U}

\newcommand{\dif}{\mathop{}\!\mathrm{d}}

\numberwithin{equation}{section}

\begin{document}
	
\markboth{T G Tenev, M F Horstemeyer}
{The Mechanics of Spacetime}

%
\catchline{}{}{}{}{}
%

\title{The Mechanics of Spacetime \textendash{} A Solid Mechanics Perspective on the Theory of General Relativity}

\author{T G Tenev\footnote{Mississippi State University, Starkville, MS 39759, USA}}

\address{Mississippi State University\\
	Starkville, MS 39759,
	USA \\
	ticho@tenev.com}

\author{M F Horstemeyer}

\address{Mississippi State University\\
	Starkville, MS 39759, USA \\
	mfhorst@me.msstate.edu}

\maketitle

\begin{history}
	\received{Day Month Year}
	\revised{Day Month Year}
\end{history}
	
\begin{abstract}
We present an elastic constitutive model of gravity where we identify physical space with the mid-hypersurface of an elastic hyperplate called the ``cosmic fabric'' and spacetime with the fabric's world volume. Using a Lagrangian formulation, we show that the fabric's behavior as derived from Hooke's Law is analogous to that of spacetime per the Field Equations of General Relativity. The study is conducted in the limit of small strains, or analogously, in the limit of weak and nearly static gravitational fields. The Fabric's Lagrangian outside of inclusions is shown to have the same form as the Einstein-Hilbert Lagrangian for free space. Properties of the fabric such as strain, stress, vibrations, and elastic moduli are related to properties of gravity and space, such as the gravitational potential, gravitational acceleration, gravitational waves, and the energy density of free space. By introducing a mechanical analogy of General Relativity, we enable the application of Solid Mechanics tools to address problems in Cosmology.


\ccode{PACS numbers:
	04.50.Kd; 
	46.90.+s} 

\end{abstract}

\section{Introduction}\label{sec:intro}

This paper explores the material analogy of physical space and its implications. Since as early as Isaac Newton~\cite{IsaacNewton1718} there have been various theories about an all pervasive cosmic medium, also known as ``ether,'' through which light and matter-matter interactions propagate. These theories culminated with the Lorentz Ether Theory (LET)~\cite{Lorentz1892,Lorentz1895}, which postulated length contractions and time dilations for objects moving through ether~\cite{FitzGerald1889,Lorentz1892} in order to explain the negative outcome of the Michelson and Morley’s ether detection experiment~\cite{Michelson1887}. Although the theory of Special Relativity (SR)~\cite{Einstein1905} appeared to obviate the need for an ether, in reality SR and LET are mathematically equivalent and experimentally indistinguishable from one another. After the development of General Relativity~\cite{Einstein1916}, which attributed measurable intrinsic curvature to spacetime, Einstein conceded~\cite{Einstein1922} that some notion of an ether must remain. More recently, theoretical predictions from Quantum Field Theory~\cite{Rugh2002} include zero-point energy density of vacuum, which further supports the material view of physical space.

The possible confluence between Hooke's Law and Einstein's Gravitational Law motivated the material analogy of space that is explored here. In 1678, Robert Hooke, a contemporary of Isaac Newton, published what later became known as Hooke's Law~\cite{Hooke1678}. In 1827, Cauchy~\cite{A.L.Cauchy1827} advanced Hooke's Law by defining the tensorial formulation of stress. For an isotropic linear elastic material, Hooke's Law states in tensorial form that,
\begin{equation}\label{eq:hookes-law}
  \sigma^{kl} = \frac{Y}{1+\nu} \left(\frac{\nu}{1-2\nu} g^{ij}g^{kl} + g^{ik}g^{jl}\right)\varepsilon_{ij}
\end{equation}
where $\sigma^{kl}$, $\varepsilon_{ij}$, and $g^{ij}$ are the stress, strain, and the metric tensors, respectively, $Y$ is Young's elastic modulus, and $\nu$ is the Poisson's ratio. Latin indexes, $i,j,k,l = {1 \ldots 3}$, run over the three spatial dimensions, and Einstein summation convention is employed. In 1916 Einstein published the field equations of General Relativity~\cite{Einstein1916}, which can be written as,
\begin{equation}\label{eq:gravity-law}
  T_{\mu\nu} = \frac{1}{\kappa} \left(R_{\mu\nu} - \frac{1}{2}R g_{\mu\nu} \right)
\end{equation}
where $T_{\mu\nu}$, $R_{\mu\nu}$, and $g_{\mu\nu}$ are the stress-energy tensor, Ricci curvature tensor, and spacetime metric tensor, respectively; $R \equiv R^\mu_\mu$ is the Ricci scalar, $\kappa \equiv 8\pi G/c^4$ is the Einstein constant as $c$ and $G$ are the speed of light and gravitational constant, respectively. Greek indexes, $\mu,\nu = 0 \ldots 3$, run over the four dimensions of spacetime with the $0^{\text{th}}$ dimension representing time. For the purposes of this paper, we have omitted the Cosmological Constant, which is sometimes included in \Eref{eq:gravity-law}, because its value is negligible for length-scales below the size of the observable universe. Einstein's Gravitational Law \eqref{eq:gravity-law} suggests a material-like constitutive relation, similar to Hooke's Law \eqref{eq:hookes-law}, because it relates stress, on the left-hand side, to deformation on the right-hand side. At first glance, the similarity appears imperfect because the right-hand sides differ in dimensionality: whereas the strain term, $\varepsilon_{ij}$, is dimensionless, the curvature terms, $R_{\mu\nu}$ and $R$, have dimensions of $\text{Length}^{-2}$. However, we resolve this problem by considering bending deformation instead of just straightforward stretch, contraction, or shear deformation. In the equations for bending, stress is proportional to the second spatial derivative of strain. 

After Einstein's publication of General Relativity~\cite{Einstein1916}, a number of researchers have investigated the relationship between Mechanics notions and General Relativity. 

One category of publications dealt with generalizing the equations of Solid Mechanics to account for relativistic effects. Synge~\cite{Synge1959} formulated a constitutive relationship in relativistic settings. Rayner~\cite{Rayner1963} extended Hooke's Law to a relativistic context. Maugin~\cite{Maugin1971} generalized the special relativistic continuum mechanics theory developed by Grot and Eringen~\cite{Grot1966} to a general relativistic context. More recently, Kijowski and Magli~\cite{Kijowski1994} presented the relativistic elasticity theory as a gauge theory. A detailed review of relativistic elasticity can be found in Karlovini and Samuelsson~\cite{Karlovini2002}.

Another category of publications interprets General Relativity in Solid Mechanical terms. Kondo~\cite{Kondo1964} mentions an analogy between the variation formalism of his theory of global plasticity and General Relativity. Gerlach and Scott~\cite{Gerlach1986} introduce a ``metric elasticity'' tensor in addition to the elasticity of matter itself and ``stresses due to geometry.'' However, these stress and strain terms are not a constitutive model of gravity, because they are not expected to apply in the absence of ordinary matter. Tartaglia~\cite{Tartaglia1995} attempted to describe spacetime as a four-dimensional elastic medium in which one of the spatial dimensions has been converted into a time dimension by assuming a uniaxial strain. However, many of the ideas in Tartaglia's paper appear to be incomplete. Antoci and Mihich~\cite{Antoci1999} explored the physical meaning of the straightforward formal extension of Hooke's Law to spacetime, but did not consider the possibility, which is explored in this paper, that Einstein's Gravitational Law may be related to Hooke's Law. Padmanabhan~\cite{PADMANABHAN2004} treated spacetime as an elastic solid and used entropy consideration to arrive at the Field Equations~\eqref{eq:gravity-law}, but unlike the work presented here, he was not concerned with developing the correspondence between the gravitational properties of cosmic space and the mechanical attributes of said solid, such as its strain and elastic modulus. Beau~\cite{Beau2015} pushed the material analogy further by interpreting the cosmological constant $\Lambda$ as related to a kind of a spacetime bulk modulus, but the analogy is to a fluid-like material and not a solid. A set of recent publications, for which Rangamani~\cite{Rangamani2009} presents a literature review, explore the applicability of the Navier-Stokes equations of Fluid Dynamics to gravity. While a fluid analogy is useful for some applications, it does not account for shear waves in space, such as gravity waves, because fluids are only capable of propagating pressure waves and not shear waves. In contrast to the prior literature recounted above, the work presented here begins with the premise that space exhibits material-like behavior subject to a constitutive relationship that can be expressed in terms of Hooke's Law~\eqref{eq:hookes-law}. 

Very recently, Hehl and Kiefer \cite{Hehl2018} related the three dimensional DeWitt Metric~\cite{Dewitt1967} with Hooke’s Law~\cite{Hooke1678} using the elastic constants as a fourth rank tensor, which was based on the work of Marsden and Hughes~\cite{Marsden1983}. Also, in relation to dark matter, B{\"{o}}hmer et al.~\cite{Bohmer2018} modified the General Theory of Relativity by using anisotropic continuum mechanics. In agreement with our results, but derived via independent means,  Hehl and Kiefer's determined that the first and second Lam\'e parameters of the cosmic material must be $\lambda = -1$ and $\mu = \frac{1}{2}$ respectively, for some appropriately chosen units, which is equivalent to stating that the Poisson ratio $\nu$ is unity, since $\nu = \frac{\lambda}{2(\lambda -\mu)}$. Nevertheless, a critical insight of our work, which has not been mentioned in any publication so far, is relating Einstein's Gravitational Law~\eqref{eq:gravity-law} to the bending deformation of a material plate as opposed to straightforward longitudinal or shear type of deformation. Without this insight, the detailed correspondence between physical space and a material medium remains largely obscured, thus limiting the practical applications of the material analogy. 

\begin{figure}
	\begin{center}
		\includegraphics[width=0.9\linewidth]{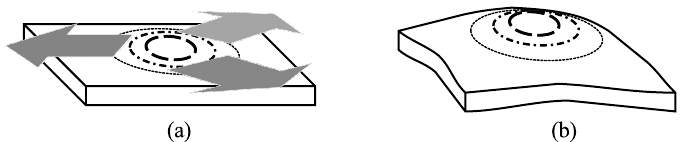}
	\end{center}
	\caption{A plate bending from flat geometry (a) into a curved geometry (b) because of an inclusion that prescribes uneven strain field, as indicated by the concentric dashed lines and the diverging arrows. The strain is larger near the center and tapers off with the distance from it. For the geometry of the plate to accommodate the prescribed strain, the plate must bend into the transverse dimension.}
	\label{fig:plate-bending}
\end{figure}

In this paper, we develop a formal analogy between Solid Mechanics and General Relativity by identifying physical space with the mid-hypersurface of a four dimensional hyperplate, called the ``cosmic fabric,'' which has a small thickness along a fourth spatial dimension and exhibits a constitutive stress-strain behavior. Matter-energy fields act as inclusions within the fabric causing it to expand longitudinally and consequently to bend. The effect, illustrated on \Fref{fig:plate-bending}, is analogous to the result from General Relativity in which matter causes space to bend resulting in gravity. Unlike other theoretical paradigms that introduce additional spatial dimensions, such as string theory~\cite{Gross1985}, or Brane world quantum models~\cite{Duff1988}, our formulation is based on conventional Solid Mechanics theories that operate strictly within the three ordinary spatial dimensions. 

We conduct our study in the limit of weak and nearly static gravitational fields, and demonstrate that outside of inclusions, the fabric's action $\SF$, assumes the form of the Einstein-Hilbert action $\SEH$,
\begin{equation}\label{eq:action}
\SF = \frac{YL}{48} \int R\sqrt{|g|}\dif x^4\quad
\text{vs.}\quad
\SEH= \frac{1}{2\kappa} \int R\sqrt{|g|}\dif x^4\; 
\end{equation}
where L is the reference thickness of the fabric, $g \equiv \det g_{\mu\nu}$, and the integral is taken over a large enough volume of spacetime sufficient to ensure convergence. The action integral of any physical system fully determines its dynamics, because the system's equations of motion can be derived from the variation of the action with respect to the metric. Therefore, once we recognize $\SF$ as analogous to $\SEH$, we can interpret various attributes of the cosmic fabric, such as its shape, strain, vibrations, and elastic moduli as analogous to properties of gravity and space, such as curvature, gravitational potential, gravitational waves, and the zero point energy density of space. 

Our approach ostensibly resembles the Arnowitt-Deser-Misner (ADM)~\cite{Arnowitt2004} and DeWitt~\cite{Dewitt1967} formulations of gravity in the way time lapse is separated from spatial extent. For example, under the ADM approach, spacetime is foliated into space-like hypersurfaces related to each other via shift and lapse functions. Like ADM, DeWitt also considers the time evolution of the three dimensional spatial metric. Nevertheless, the Cosmic Fabric model differs from  these formulations in that it associates constitutive behavior with the geometric description of gravity and derives its governing equations from a material-like constitutive relation. Furthermore, unlike these formulations, the cosmic fabric specifies a hyperplane of absolute simultaneity. 

The Cosmic Fabric model of gravity allows General Relativity problems to be formulated as Solid Mechanics problems, solved within the Solid Mechanics domain, and the solution interpreted back in General Relativity terms. The reverse is also true. Thus, ideas, methodologies and tools from each field become available to the other field. Over the past century, Solid Mechanics and General Relativity have advanced independently from each other with few researchers having expertise in both. Consequently, significant terminology and focus gaps exist between these two fields, which obscure their underlying physical similarities. Our research attempts to bridge these gaps.

The remainder of this paper is organized as follows: In \Sref{sec:formulation} we develop the Solid Mechanics analogy of gravity by specifying a material body whose behavior, determined solely based on Hooke's Law \eqref{eq:hookes-law}, is demonstrably analogous to the behavior of spacetime. In \Sref{sec:discussion} we discuss the implications of the resulting model, and summarize and conclude in \Sref{sec:summary}.

\section{Formulation of the Cosmic Fabric Model of Gravity}\label{sec:formulation}

Consider a four dimensional hyperplate, called here the ``cosmic fabric,'' which is thin in the fourth spatial dimension, $x^4$. We show that, for a suitably chosen constitutive parameters, the fabric's Lagrangian density outside of inclusions is $\LF = (YL/48)R\sqrt{|g|}$, where $\LF$ is the integrand in \Eref{eq:action}. This result enables us to subsequently analyze how the remaining kinematic properties of the cosmic fabric correspond to properties of gravity.

\subsection{Notation}

For the remainder of this paper, we will use the following notation and conventions: Lower case Latin indexes, $i,j,k,l = 1\ldots3$ run over the three ordinary spatial dimensions. Upper case Latin indexes, $I,J,K,L = 1\ldots4$ run over the four hyperspace dimensions, while Greek indexes, $\mu,\nu,\alpha = 0\ldots3$ run over the four spacetime dimensions, where indexes $0$ and $4$ represent, respectively, the time dimension and the extra spatial dimensions. Sometimes we will use $\xi$, where $\xi \equiv x^4$, for the thin dimension of the hyperplate. Also, we will use $t$ to denote coordinate time such that $x^0 \equiv ct$, where $c$ is the speed of light. Indexes appearing after a comma represent differentiation with respect to the indexed dimension. For example, $u_{i,j} \equiv \partial u_i/\partial x^j$. For spacetime, we adopt the space-like metric signature $(-,+,+,+)$ and denote the flat metric tensor as $\eta_{\mu\nu}$, where $[\eta_{\mu\nu}] \equiv \operatorname{diag}[-1,1,1,1]$.

\subsection{Coordinate Assignment and Reference Space} \label{ssec:coords}

\begin{figure*}
	\centering
	\includegraphics[width=0.9\linewidth]{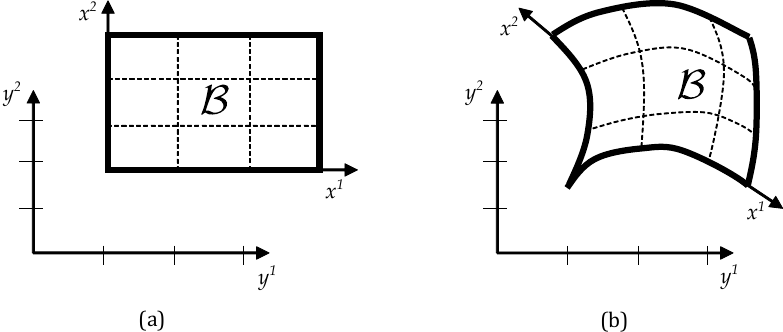}
	\caption{Material coordinates $x^I$ versus reference coordinates $y^I$, before (a) and after (b) deformation of a body $\mathcal{B}$. The material coordinates are attached to each material point and are carried along with the material as it deforms, while reference coordinates remain fixed during deformation. In the diagram, two of the spatial dimensions have been suppressed for clarity.}
	\label{fig:material-ref-coords}
\end{figure*}

We consider the cosmic fabric as immersed in a four dimensional (4D) hyperspace within which it can deform. Moreover this 4D hyperspace is flat and has been assigned a Cartesian coordinate grid with coordinates $y^I$. Another set of coordinates $x^I$ is painted on the fabric in the following manner: prior to deformation the $x^I$ coordinates are painted such that they coincide with the $y^I$ coordinates.  As the fabric deforms (see \Fref{fig:material-ref-coords}), the $x^I$ coordinates remain attached to each material point and displace along with it. We call $y^I$ the \emph{reference coordinates}, and we call the 4D hyperspace the \emph{reference space}. Also, we call $x^I$ the \emph{material coordinates}, because they name material points. At any given moment, each point on the fabric can be specified by either its reference coordinates $y^I$ or material coordinates $x^I$, such that prior to deformation, $y^I = x^I$. These two sets of coordinates are commonly used in Solid Mechanics where "reference coordinates" are also known as "spatial coordinates." The reference space described here is a mathematical construct that helps us build the analogy between Solid Mechanics and General Relativity, but unlike the cosmic fabric itself, it is not necessarily a physical entity. For example, an observer within the fabric is unable to measure directly any attributes of this reference space. 

With respect to the reference space, metric rulers do not change length as the fabric deforms. When the fabric is stretched the number of rulers that can fit between two given points increases. Since metric rulers define the unit of length within the fabric, its stretching is perceived from within the fabric as the expansion of physical space. The term \emph{strain} refers to either the stretch or contraction of a body. The differential straining of the fabric gives rise to its intrinsic curvature and is perceived from within it as the intrinsic curvature of physical space. 

The fabric and its enclosing reference space share the same coordinate time $t$. Thus, their respective time coordinates $x^0$ and $y^0$ are such that $x^0 = y^0 \equiv ct$. Note, however, that the proper time $\tau$, which is measured by clocks within with the fabric, is not necessarily one and the same as $t$, and in general, $d\tau/dt \le 1$.

\subsection{Deformation Basics}\label{ssec:basics}

Let $x^i$ be the material coordinates assigned to the cosmic fabric's mid-hypersurface, and let $g_{ij}$ be the metric tensor of the fabric. The metric tensor defines how coordinate differences relate to distances. Thus, the distance $ds$ between two nearby material points is given by, 
\begin{equation}
ds^2 = g_{ij}dx^i dx^j
\end{equation}
For Cartesian coordinates, as adopted here, the distance $d\overbar{s}$ between the same two material points prior to deformation is,
\begin{equation}
d\overbar{s}^2 = \delta_{ij}dx^i dx^j
\end{equation}
where $\delta_{ij}$ is the Kronecker delta.

The strain tensor $\e_{ij}$ quantifies the amount of relative length change during deformation. By definition, $\e_{ij}$ is such that,
\begin{equation}\label{eq:strain}
\begin{split}
2\e_{ij}dx^i dx^j & = ds^2 - d\overbar{s}^2 = (g_{ij} - \delta_{ij})dx^i dx^j \\
\therefore \e_{ij} &= \frac{1}{2}(g_{ij} - \delta_{ij})
\end{split}
\end{equation}

\begin{figure*}
	\centering
	\includegraphics[width=0.9\linewidth]{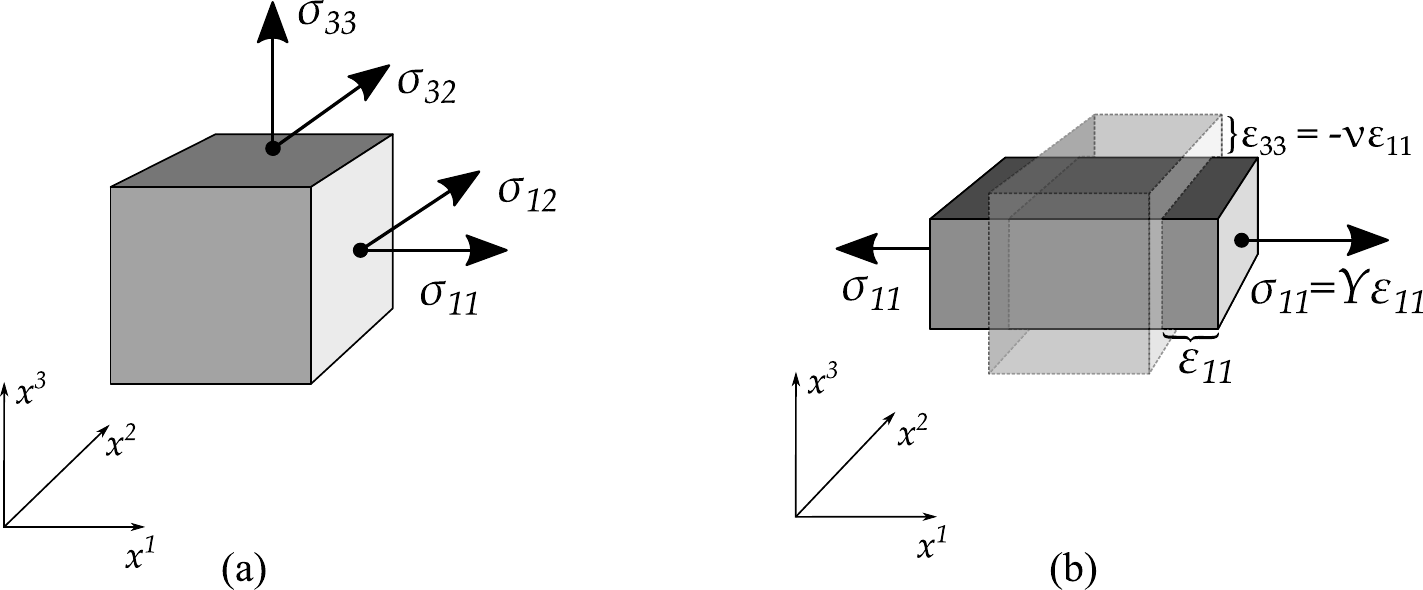}
	\caption{Multi-axial stress state (a), and a uniaxial deformation of an object (b) from the transparent to the deformed shape. Each component $\sigma_{ij}$ represents the stress through the $i^\text{th}$ surface in the $j^\text{th}$ direction. The Poisson's ratio $\nu$ measures the effect of the longitudinal stress along the $i^\text{th}$ direction on the longitudinal strain along the $j^\text{th}$ direction, for $j \neq i$. In the case of uniaxial stress state, $\varepsilon_{jj} = (-\nu/Y) \sigma_{ii} = -\nu\varepsilon_{ii}$.}
	\label{fig:stress-strain-poisson}
\end{figure*}

The 3D volumetric strain, defined as
\begin{equation}
\ethree \equiv \e^i_i
\end{equation}
is a scalar field that represents the fractional increase of the fabric's mid-hypersurface volume. In other words, $dV/d\overbar{V} = (1+\ethree)e$, where $dV$ and $d\overbar{V}$ are, respectively the deformed and undeformed volume elements.

The Young's elastic modulus $Y$, which figures in Hooke's Law \eqref{eq:hookes-law}, is the amount of longitudinal stress (force per unit of cross section area) $\sigma_{ii}$ in the $i^\text{th}$ direction needed to produce a unit amount of longitudinal strain $\varepsilon_{ii}$ in the same direction under a uniaxial stress condition (no summation intended over the index $i$). The effect of longitudinal stress along a given orientation on the longitudinal strains in the transverse orientations is known as the Poisson effect and is measured by the Poisson's ratio $\nu$ (see \Fref{fig:stress-strain-poisson}).

\subsection{Postulates} \label{ssec:postulates}

We postulate the cosmic fabric to be (1) an elastic thin hyperplate, with (2) matter-energy fields as inclusions, and (3) lapse rate of proper time proportional to the shear wave speed $v_s$. Each of these postulates is described and motivated in the sections below.

\subsubsection{Elastic Thin Hyperplate} \label{ssec:hyperplate}

Cosmic space is identified with the mid-hypersurface of a hyperplate called the Cosmic Fabric that is thin along the fourth spatial dimension. We imagine the fabric as foliated into 3D hypersurfaces each of which is isotropic and elastic, and each is subject to Hooke's Law (See \Fref{fig:hypersurfaces}). Thus, Hooke's Law \eqref{eq:hookes-law} together with concepts such as stress, strain and the Poisson effect (see \Fref{fig:stress-strain-poisson}) apply as conventionally understood in Solid Mechanics, because they pertain to individual hypersurfaces, which are 3D bodies.

Because of its correspondence to physical space, the intrinsic curvature, $\Rthree$, of the fabric's mid-hypersurface corresponds to that of three-dimensional (3D) space. Likewise, the intrinsic curvature $R$ of the fabric's world volume, corresponds to that of four-dimensional (4D) spacetime. The term ``world volume'' refers to the four-dimensional shape traced out by an object in spacetime as it advances in time.

The small transverse thickness of the fabric is needed to create resistance to bending, but once such resistance is accounted for, we treat the fabric as essentially a 3D hypersurface that bends within the 4D reference hyperspace. The thickness must be very small so that the fabric can behave as an essentially 3D object at ordinary length scales and be an appropriate analogy of 3D physical space. The thickness itself defines a microscopic length scale at which the behavior of the physical world would have to differ significantly from our ordinary experience. A value equal or comparable to Planck's length $l_p$ meets this criteria. However, the exact value of the thickness is not essential to the model as long as it is small but not vanishingly so.
	
\subsubsection{Inclusions} \label{ssec:inclusions}

Matter-energy fields behave as inclusions in the fabric inducing \emph{membrane} strains leading to transverse displacements and hence bending (\Fref{fig:plate-bending}). The following equation postulates that matter is a source of volumetric strain,
\begin{equation}\label{eq:strain-laplacian}
\ethree_{,kk} \propto c^2\kappa \rho
\end{equation}
where $\ethree_{,kk} \equiv \nabla^2\ethree$ is the Laplacian of the volumetric strain, $c$ is the speed of light, $\kappa$ is the Einstein constant, and $\rho$ is the density of matter-energy. The term ``membrane'' strain (or stress) refers to strains (or stresses) that change in-plane but are uniform across the thickness of the fabric as opposed to bending strains (or stresses) that switch sign through the thickness across the mid-hypersurface.

The mass content of matter, rather than its spatial extent, is what causes the displacement of fabric material. In the context of General Relativity, mass can be related to geometry through its Schwarzchild radius. Thus, one meter of mass is the amount of mass whose Schwarzchild radius is two meters. In the same way, the geometric significance of a matter-energy field, represented by the right hand side of \Eref{eq:strain-laplacian}, can be understood as the Schwarzchild radius density and $c^2\kappa$ as a units conversion factor. In other words, \Eref{eq:strain-laplacian} postulates that the Schwarzchild radius density of a matter-energy field is a source of volumetric strain in the cosmic fabric.

The analogy between a body in empty space and an inclusion in the cosmic fabric raises the question of how such an inclusion can move freely through a stiff fabric in the same way as a body can move through empty space. The wave nature of matter, at the length-scale of the body's elementary particles suggests the answer. Just like waves can propagate through a very stiff material, in the same way, elementary particles, which have wave nature, could propagate through the fabric. A detailed treatment of the matter-fabric interaction requires extending our model with a mechanical analogy for the nature of matter, which is beyond the scope of this paper. Instead, the details of the underlying matter-fabric interactions are abstracted and only the effect is considered, namely, that matter inclusions prescribe a strain field on the fabric. This strain field is then treated as the input to our model. Representing matter as a strain field within the fabric allows us to aggregate the effects of individual elementary particles over large length scales, and treat planets and stars as individual inclusions.

\subsubsection{Lapse Rate}

The Lapse Rate postulate relates the flow of proper time to the geometry of the cosmic fabric. All matter-matter interactions are mediated by signals propagating in the fabric as shear waves. Therefore, the rate of such interactions varies proportionally to the shear wave speed. A clock placed where fabric waves propagate slower would tick proportionally slower compared to a clock placed where fabric waves propagate faster. Such effect is independent of the clock’s design, because the speed of fabric waves affects all matter-matter interactions. In other words, the lapse rate at each point in the fabric, that is how fast clocks tick, is proportional to the speed of shear waves propagating in the fabric when measured in relation to the reference space.

Notice that the shear wave speed will appear to have remained constant when measured by an observer within the fabric, because the reduction in lapse rate exactly compensates for the reduction in shear wave speed. This perceived invariance of the shear wave speed is analogous to the speed of light invariance in General Relativity. 

Stated quantitatively, we postulate that the shear wave speed $v_s$ depends on the fabric's volumetric strain $\ethree$ as follows,
\begin{equation}\label{eq:v_s}
v_s = (1 + \ethree)^{-1}c
\end{equation}
Consequently, the lapse rate, that is the relationship between proper time $\tau$ and coordinate time $t$, is as follows, 
\begin{equation}\label{eq:tau}
\frac{d \tau}{d t} = (1 + \ethree)^{-1}
\end{equation}

We motivate the above postulate by connecting the shear wave speed $v_s$ to the mechanical properties of the cosmic fabric. 
A well known result from Solid Mechanics is that $v_s = \sqrt{\mu/\rho}$ where $\mu$ and $\rho$ are, respectively, the shear modulus and density of the material. When such material is stretched, its density decreases by a factor of $(1+\ethree)$ because the same amount of material now occupies $(1+\ethree)$ times more volume. The elastic modulus also changes when the fabric is stretched, but its relationship to strain depends on the internal structure of the material. The choice of modulus-strain relationship becomes a parameter in our model that controls the effect of time dilation. By fixing this relationship to be such that,
\begin{equation}\label{eq:mu}
\mu = (1+\ethree)^{-3}\mu_0
\end{equation}
where $\mu_0$ is the reference modulus of the undeformed fabric, we can recover \Eref{eq:v_s}. One reason why the modulus changes is that the internal structure of the material weakens under stretch. As discussed in \Cref{Allison2012}, there are materials which exhibit modulus-strain relationship similar to the one in \Eref{eq:mu}.

\subsection{Weak and Nearly Static Fields Condition} \label{ssec:simplifications}

To keep the math tractable, we conduct our study under the assumption of weak and nearly static fields. We believe that this assumption is not fundamental to the model, and that it could be relaxed or removed in the future. As will be shown in \Sref{ssec:strain}, the fabric strain is analogous to the gravitational potential, so the weak field condition, which is the subject of Linearized Gravity~\cite{Misner1973Ch18} is analogous to the small strain condition, which is the subject of Solid Mechanic's Infinitesimal Strain Theory. 

We consider a gravitational potential $\Phi$ to be weak if $\left| \Phi / c^2 \right| \ll 1$. By this definition, most gravitational fields that we experience on an everyday basis are weak. For example, the values for $\left| \Phi / c^2 \right|$ at the Earth's surface due to the gravitational fields of the Earth, Sun, and Milky way are $6.7\times 10^{-10}$, $1.0\times 10^{-8}$, and $1.4 \times 10^{-6}$, respectively~\cite{WorldHeritageEncyclopedia2014}.  As such, we consider these gravity fields to be weak.

Except in regards to gravity waves (\Sref{ssec:wave-equation}), we will assume nearly static fields (or slow changing) in addition to weak fields. A field is considered nearly static if the gravitating masses causing the field to move with velocities are much less than the speed of light. This is the case for most gravitational fields that we experience. The nearly static field assumption means that differentiation with respect to time results in negligibly small values.

\subsection{Linearized Spacetime Metric}

Under the weak field condition, the metric tensor can be approximated as,
\begin{equation}\label{eq:weak-field-g}
g_{\mu\nu} = \eta_{\mu\nu} + 2\varepsilon_{\mu\nu}, \quad |\varepsilon_{\mu\nu}| \ll 1 
\end{equation}
where the term $2\e_{\mu\nu}$ plays the same role as the small quantities $h_{\mu\nu}$ that are commonly used in General Relativity literature in discussions on Linearized Gravity, such as in \Cref{Misner1973Ch18}. However, note that except under special conditions, $\e_{\mu\nu}$ does not necessarily comply with the harmonic gauge condition, which is often employed in Linearized Gravity.

From \eref{eq:strain} we recognize the spatial components $\e_{ij}$ as the strain of the fabric's mid-hypersurface. The component $\e_{00}$, as well as the other time components, are related to the flow of proper time. Below, we compute a relationship between $\e_{00}$ and the fabric's strain.

From \eref{eq:tau} we can deduce an expression for the time-time component $g_{00}$ of the fabric's spacetime metric $g_{\mu\nu}$ as follows. Applying the metric equation for a stationary point on the fabric,
\begin{equation}\label{eq:g00}
\begin{split}
-c^2 d\tau^2 &= g_{00} c^2 dt^2 = -(1+\ethree)^{-2} c^2 dt^2  \\
\therefore g_{00} &= -(1+\ethree)^{-2}
\end{split}
\end{equation}
Combining Equations \eqref{eq:g00} and \eqref{eq:weak-field-g},  we note that
\begin{equation}\label{eq:e00}
\begin{split}
-1 + 2\e_{00}  &\approx g_{00} = -(1+\ethree)^{-2} \approx -1 + 2\ethree \\
\therefore & \e_{00} = \ethree
\end{split}
\end{equation}
which, by the application of the Inclusion Postulate \eqref{eq:strain-laplacian}, yields the following result,
\begin{equation} \label{eq:e00-as-ekk}
\e_{00,kk} = \ethree_{,kk} \propto c^2 \kappa \rho
\end{equation}

\subsection{Bending Energy Density}\label{ssec:bending-energy}

Rather than treating the fabric as a 4D hyperplate, it is convenient to approximate it as a 3D hypersurface. This can be accomplished once we have averaged the fabric's elastic energy density $\calU$ across its thickness and assign it to its mid-hypersurface. At that point, we can use the fabric's mid-hypersurface as a proxy instead of the fabric in future calculations. 

\begin{figure*}
	\centering
	\includegraphics[width=.9\linewidth]{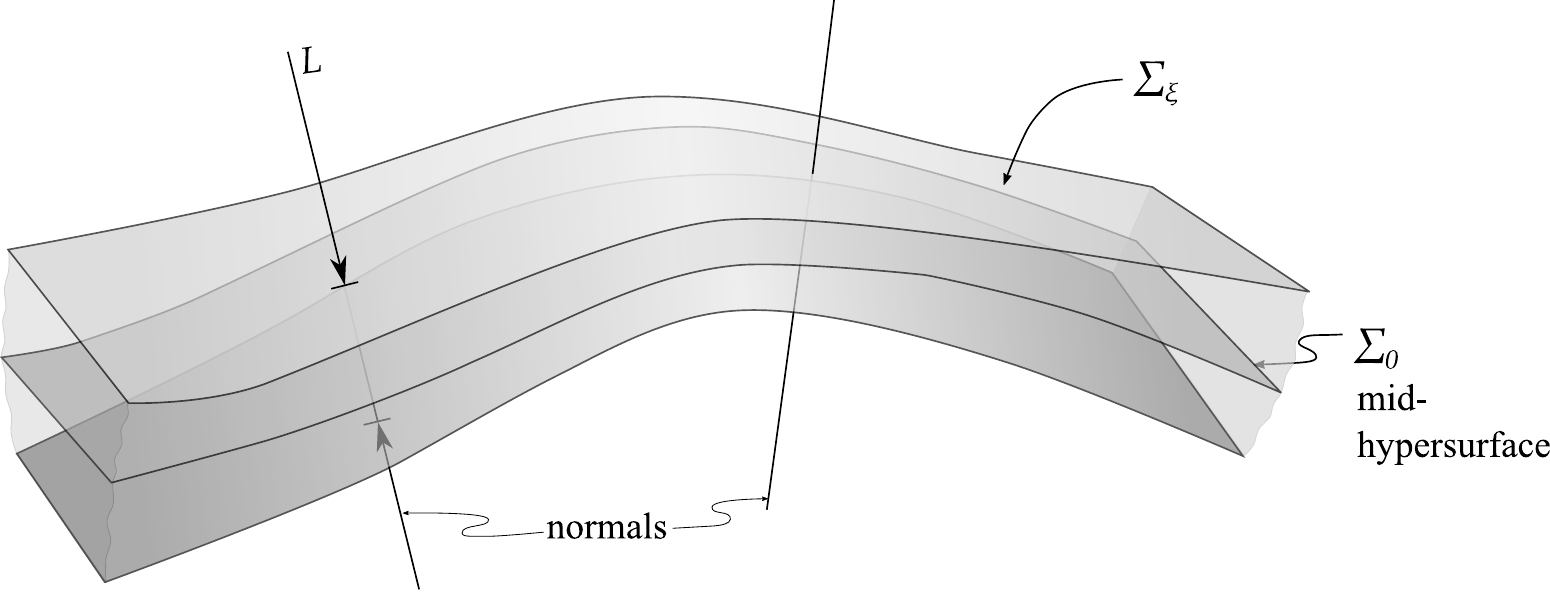}
	\caption{The cosmic fabric is treated as a stack of three-dimensional hypersurfaces $\Sigma_{\xi}$ each parameterized by $\xi \equiv x^4 = \text{const}$, and $L$ is its thickness.}
	\label{fig:hypersurfaces}
\end{figure*}

To compute $\calU$, we adapt the work of Efrati \emph{et al.}~\cite{Efrati2008} concerning the bending of conventional thin plates. For ease of notation, let $\xi \equiv x^4$ denote the coordinate offset from the mid-hypersurface of the fabric. The fabric, is regarded as foliated into infinitely many hypersurfaces $\Sigma_\xi$ each parameterized by $\xi = \text{const.}$ (\Fref{fig:hypersurfaces}). We carry over the simplifying assumption from Kirchoff-Love thin plate theory~\cite{Love1888} to thin hyperplates and assume that the set of material points along any given hypersurface that were along a normal prior to bending remain along the normal after bending. 

It can be shown \cite{Efrati2008} that the metric $g_{ij} = g_{ij}(\xi)$ of each $\Sigma_{\xi}$ takes the form,
\begin{equation}\label{eq:g-hypersurface}
g_{ij} = a_{ij} - 2b_{ij} \xi + c_{ij} \xi^2
\end{equation}
where $a_{ij} = a_{ij}(x^i)$ and $b_{ij} = b_{ij}(x^i)$ are, respectively, the first and second fundamental forms of the mid-hypersurface, and $c_{ij} = a^{kl}b_{ik}b_{jl}$. 

The total elastic energy density of a linearly elastic solid is half of the inner product of its stress and strain tensors. 
The dependence of the cosmic fabric's modulus on strain adds a degree of nonlinearity, which would have resulted in a correction factor of about $(1+\ethree)^{-2}$. However, under the small strain conditions, $\ethree \ll 1$, allowing us to approximate $(1+\ethree)^{-2} \approx 1$, and thus to neglect the nonlinear effect. Applying Hooke's Law \eqref{eq:hookes-law}, the total elastic energy density $\calU_{\xi}$ of each hypersurface $\Sigma_{\xi}$ is given by,
\begin{equation}\label{eq:elastc-energy}
\begin{split}
\calU_{\xi} &= \frac{1}{2} \sigma^{ij} \varepsilon_{ij} 
= \frac{1}{2} C^{ijkl}\varepsilon_{ij}\varepsilon_{kl} \\
& \text{such that} \quad C^{ijkl} \equiv \frac{Y}{1+\nu} \left(\frac{\nu}{1-2\nu} g^{ij}g^{kl} + g^{ik}g^{jl}\right)
\end{split}
\end{equation}
where $\sigma_{ij} = \sigma_{ij}{(\xi)}$ and $\varepsilon_{ij} = \varepsilon_{ij}(\xi)$ are, respectively, the stress and strain at each hypersurface $\Sigma_{\xi}$. Note that here and and for the remainder of the paper, we compute the elastic energy density with respect to the coordinate volume as opposed to the proper volume.

Next, we compute the total elastic energy density $\calU$ averaged across the fabric's thickness, and we separate it into a bending term $\UB$ and a membrane stretch term $\UM$. For this purpose, we split the strain at each surface, $\varepsilon_{ij}$ into a membrane strain $\eM_{ij}$ and a bending strain $\eB_{ij}$ as follows:
\begin{equation}
\begin{split}
\varepsilon_{ij}  & = \frac{1}{2}(g_{ij} - \delta_{ij}) = \eM_{ij} + \eB_{ij} \\
\eM_{ij} & = \frac{1}{2}(a_{ij} - \delta_{ij}) \\
\eB_{ij} & = -b_{ij}\xi   + \calO (b^2\xi^2)
\end{split}
\end{equation}

\begin{equation}\label{eq:energies}
\begin{split}
\mathcal{U} & = \frac{1}{L} \int_{-\frac{L}{2}}^{\frac{L}{2}} \calU_{\xi} \dif \xi \\
& = \frac{1}{2L} \int_{-\frac{L}{2}}^{\frac{L}{2}} C^{ijkl}(\eM_{ij}\eM_{kl} + \eB_{ij}\eB_{kl} + [\eM_{ij}\eB_{kl} + \eB_{ij}\eM_{kl}]) \dif \xi \\
& = \UM + \UB \\
\UM &= \frac{1}{2L} \int_{-\frac{L}{2}}^{\frac{L}{2}} C^{ijkl}\eM_{ij}\eM_{kl}\dif \xi \\
\UB &= \frac{1}{2L} \int_{-\frac{L}{2}}^{\frac{L}{2}} C^{ijkl}\eB_{ij}\eB_{kl}\dif \xi \\
\end{split}
\end{equation}
The term $\calO(b^2\xi^2)$ stands for an expression whose order of magnitude is comparable to the squire of the elements $b_{ij}$ multiplied by $\xi^2$. The mixed terms inside the square brackets in \Eref{eq:energies} vanish under integration because the bending strain reverses sign across the mid-hypersurface; hence $\eB_{ij} = \eB_{ij}(\xi)$ is an odd function, while $\eM_{ij} = \eM_{ij}(\xi)$ is an even function.

For the remainder of this subsection, we focus on evaluating the term $\UB$. The term $\UM$ will be addressed in the following subsection where we show that it vanishes under appropriately chosen material properties and deformation kinematics.

Evaluating $\UB$ from \Eref{eq:energies}, we obtain,
\begin{equation}
\UB = L^2C^{ijkl}\left[b_{ij}b_{kl} + \calO(b^3 L)\right]
\end{equation}
The extrinsic curvature terms $b_{ij}$ have magnitudes comparable to the inverse of the curvature radius. The curvature radius is much greater than the thickness of the fabric, so $\calO(b L) \ll 1$ allows us to neglect the term $\calO(b^3 L)$ in the above expression. Using the identity, $\Rthree_{lijk} = b_{ik}b_{jl} - b_{ij}b_{kl}$, where $\Rthree_{lijk}$ is the Riemann curvature tensor of the mid-hypersurface, and setting $\calO(b^3 L) = 0$, we can  express $\UB$ in terms of the intrinsic three-dimensional spatial curvature $\Rthree$ as follows,
\begin{equation} \label{eq:U-general}
\UB = -\frac{L^2 Y}{24(1+\nu)} \left(
\Rthree +
\frac{1-\nu}{1-2\nu} b^i_i b^k_k 
\right)
\end{equation}

The Poisson's ratio of the cosmic fabric had remained unspecified as a freedom to be fixed at a later time such as now. In order for $\UB$ to be physical, it should not depend on the extrinsic curvature $b_{ij}$ that is not already incorporated into the intrinsic curvature $\Rthree$. The $b^i_i b^k_k$ term would vanish if we chose Poisson's ratio $\nu = 1$. In this case, the bending energy becomes as follows,
\begin{equation} \label{eq:U}
\UB = - \frac{YL^2}{48} \Rthree
\end{equation}
subject to the condition,
\begin{equation}\label{eq:mat-conditions}
\nu = 1
\end{equation}

\subsection{Membrane Energy Density}\label{ssec:membrane-energy}

We now show that for any given small-strain deformed configuration, we can identify a material displacement field that results in no membrane energy. Consequently, we conclude that the bending energy $\UB$ is the only contribution to the total elastic energy of the fabric for the case of nearly static fields. Since General Relativity (GR) is only concerned with the curvature of the deformed body, in developing the material analogy of GR we have freedom to prescribe a specific material displacement field for the deformation.

Let us consider a displacement field where each point of the mid-hypersurface, $x^4 = 0$, of the fabric is displaced within a reference space by the amount $w = w(x^i)$ along a geodesic normal to the  hypersurface. It should be evident that using such a displacement field, one can deform a flat body into any given shape that represents a small deviation from flatness and does not contain folds. Let $y^I$ be the coordinates in reference space of the position to which the material point at $x^i$ is displaced. Thus, $y^i = x^i$ and $y^4 = w$. The metric tensor of the deformed hypersurface can be computed from the dot product of the position differentials as follows,
\begin{equation} \label{eq:push-strain}
\begin{split}
g_{ij} & = y^K_{,i}y^K_{,j} = x^k_{,i}x^k_{,j} + w_{,i}w_{,j} = \delta_{ij} + w_{,i}w_{,j} \\
\therefore \varepsilon_{ij} & = \frac{1}{2}(g_{ij} - \delta_{ij}) = \frac{1}{2}  w_{,i}w_{,j}
\end{split}
\end{equation}
Using the formula for elastic energy density, $\UM = \sigma^{kl}\varepsilon_{kl}/2$ and applying Hooke's Law \eqref{eq:hookes-law} to \Eref{eq:push-strain} with $\nu = 1$, we find,
\begin{equation}\label{eq:UM}
\begin{split}
\UM & \propto \sigma^{kl}\varepsilon_{kl} \propto (g^{ik}g^{jl} - g^{ij}g^{kl})\varepsilon_{ij}\varepsilon_{kl} = \\
& = \varepsilon^k_j \varepsilon^j_k - \varepsilon^j_j \varepsilon^k_k 
\propto w^{,k}w_{,j}w^{,j}w_{,k} - w^{,j}w_{,j}w^{,k}w_{,k} = 0 \\
\therefore \UM & = 0
\end{split}
\end{equation}

Hence, fixing the fabric's deformation to material displacements only along the hypersurface normals is a valid approximation under the assumption of nearly static fields. In such cases, the reason for the deformation would have been to geometrically accommodate inclusions by bending into the $y^4$ dimension. Once bending has taken place, the material points of the fabric can shift within the plane of the fabric to minimize its membrane energy without affecting the geometrical constraints imposed by the inclusion. 
For nearly static situations, we have shown that the membrane energy can be minimized to where it vanishes. In such cases, the net displacement would have taken the form described in this subsection. 

\subsection{Lagrangian Density}\label{ssec:lagrangian-density}

Ignoring the kinetic energy component, under the simplifying assumption of nearly static fields, the Lagrangian density is $\LF = - \UB \sqrt{|g|} \propto \Rthree \sqrt{|g|}$, where $g \equiv det[g_{\mu\nu}]$. The factor $\sqrt{|g|}$, which converts from a coordinate volume to a proper volume, is needed for $\LF$ to be a tensor density, which requires invariance under coordinate transformations. 

Next, we derive an expression for $\LF$ in terms of the Ricci curvature $R$ of the fabric's world volume. According to the gauge-invariant linearized expression for $R$ per \Cref{Misner1973Ch18},
\begin{equation}\label{eq:ricci}
\begin{split}
R & = 2\left( -\e_{\mu\;,\alpha}^{\;\mu\;\;\alpha} + \e_{\;\;\;,\alpha\mu}^{\alpha\mu} \right) \\
& = 2\left( -\e_{i\;,k}^{\;i\;\;k} + \e_{\;\;\;,ik}^{ik} - \e_{0\;,k}^{\;0\;\;k} -  \e_{k\;,0}^{\;k\;\;0} + 2\e_{0k,}^{\;\;\;0k}
\right) \\
& \approx \Rthree + 2\e_{00,kk}
\end{split}
\end{equation}
In the last step of the above derivation, we have recognized that the purely spatial terms add up to the gauge-independent linearized expression for $\Rthree$. Furthermore, the terms differentiated with respect to $x^0$ are negligible because of the nearly static fields assumption. Also, lowering or raising a single $0$ index, which is accomplished using $\eta_{\mu\nu}$, changes the sign of the term.

In free space $\e_{00,kk} = 0$ per \Eref{eq:e00-as-ekk}. Consequently, after combining Equations \eqref{eq:U} and \eqref{eq:ricci}, we finally arrive at,
\begin{equation} \label{eq:lagrangian-result}
\LF =  -\UB \sqrt{|g|} = \frac{YL^2}{48} R \sqrt{|g|}
\end{equation}
which has the same form as the Einstein-Hilbert Lagrangian density. The resulting action is simply the integral of the Lagrangian density over coordinate spacetime, namely, 
\begin{equation}
\SF = \int \LF \dif x^4 = \frac{YL}{48} \int R\sqrt{|g|}\dif x^4
\end{equation}
which is what we had set out to demonstrate as stated earlier per \Eref{eq:action}.

\section{Discussion} \label{sec:discussion}

In the previous section, we postulated a material body, which we named the ``cosmic fabric'' whose constitutive behavior outside of inclusions is analogous to the behavior of gravity, and have shown the sequential mathematical development. For the analogy to be useful, it should allow us to map between notions in Solid Mechanics and General Relativity. Such a mapping is possible on the basis of identifying the fabric Lagrangian density $\LF$ with the Lagrangian density from the Einstein-Hilbert action, $\LEH$, as applying to free space. Specifically,
\begin{equation}\label{eq:identification}
\LF = \frac{YL^2}{48}R\sqrt{|g|} = \LEH =  \frac{1}{2\kappa}R\sqrt{|g|}
\end{equation}
where $\kappa$ is the Einstein constant. 

In the subsections below, we discuss the correspondence between mechanical properties of the cosmic fabric and known properties of gravity.

\subsection{Fabric Strain and Gravitational Potential}\label{ssec:strain}

It is a well known result from Linearized Gravity that given the choice of coordinates adopted here, the classical gravitational potential $\Phi$ is related to the time-time component of the metric in the following way~\cite{Misner1973Ch18},
\begin{equation}
\Phi/c^2 = -\left(g_{00} - \eta_{00}\right)/2
\end{equation}
Combined with \Eref{eq:e00} the above becomes, 
\begin{equation}
\Phi/c^2 = -\left(g_{00} + 1\right)/2 = -\e_{00} = -\ethree
\end{equation}

In other words, the gravitational potential corresponds to the volumetric expansion of the fabric.

\subsection{Poisson's Ratio and the Substructure of Space}

Known materials with a Poisson's ratio of $\nu = 1$ have a fibrous substructure, which suggests that the cosmic fabric is, in fact, a fabric! For $\nu = 1$, the bulk modulus is $K = Y/[3(1-2\nu)] < 0$. A negative bulk modulus means that compressing the fabric results in an overall increase of the material volume and vice versa. Although such behavior is unusual for most conventional materials, there are recently discovered \emph{compressive dilatant}~\cite{Rodney2016} and \emph{stretch densifying}~\cite{Baughman2016} materials, for which $\nu=1$ in either compression or tension, respectively. Compressive dilatant materials are artificially manufactured and their substructure consists of entangled stiff wires. Stretch densifying materials, have textile-like substructure comprised of woven threads each consisting of twisted fibers.

\subsection{Fabric Vibrations and Gravitational Waves}\label{ssec:wave-equation}

Having Poisson's ratio $\nu = 1$ also implies that there can only be transverse (shear) waves in the fabric but no longitudinal (pressure) waves. The shear modulus $\mu$ and the p-wave modulus $M$ are as follows,
\begin{equation} \label{eq:moduli}
\begin{split}
\mu & = \frac{Y}{2(1+\nu )} = \frac{Y}{4} \\
M &= Y \frac{1-\nu}{(1-2\nu)(1+\nu)} = 0
\end{split}
\end{equation}
implying that the transverse (shear) wave velocity $v_s = \sqrt{\mu/\rho} \neq 0$, while the longitudinal (pressure) wave velocity $v_p = \sqrt{M/\rho} = 0$. This result shows why the speed of light is the fastest entity of the universe, given that a longitudinal wave is typically faster than a shear wave. For a shear wave to be the fastest, the Poisson's Ratio must be 1. This conclusion is consistent with observations, because all known waves that propagate in free space, such as gravity or electromagnetic waves, are transverse. 

Let us consider the analogy between shear waves in the fabric and gravitational waves. Such an analogy depends on demonstrating that the fabric's behavior parallels that of spacetime for fast changing fields as well. We leave the rigorous proof for a future article, and for the rest of this subsection we assume that the fabric's behavior implied by the Lagrangian \eqref{eq:lagrangian-result} also holds for fast changing fields. Based on this assumption, we proceed to investigate in-plane shear waves propagating through the fabric and their correspondence to gravitational waves.

First, we show that if static fields are negligible and in the absence of torsion, then the strain $\e_{\mu\nu}$ satisfies the harmonic gauge condition, $\e^{\mu\alpha}_{\;\;\;,\alpha} = (1/2)\e^{\alpha\;\;\mu}_{\;\alpha,}$. For shear waves, $\ethree = 0$, and by \Eref{eq:e00}, $e_{00} = 0$, implying that $\e^{\alpha}_{\;\alpha} = 0$. Therefore, proving the harmonic gauge condition reduces to demonstrating that, $\e^{\mu\alpha}_{\;\;,\alpha} = 0$. Furthermore, the shear time-space components must vanish, $\e_{4j} = \e_{j4} = 0 = \e_{0j} = \e_{j0}$,  because we are assuming negligible static fields and in-plane shear waves. Therefore, in order to prove that the harmonic gauge condition holds, we just need to show that $\e_{ik,k} = 0$. Let $u_i$ be the material displacement field. In terms of the displacement field, the strain is $2\e_{ij} = u_{i,j} + u_{j,i}$, and so,
\begin{equation}
\begin{split}
2\e_{ij} &= 2u_{j,i} + [u_{i,j} - u_{j,i}]   \\
2\e_{ik,k} &= 2u_{k,ki} + [u_{i,k} - u_{k,i}]_{,k}
\end{split}
\end{equation}
But, $u_{k,ki} = 0$ since $\e_{kk} = u_{k,k} = 0$. The difference in the square brackets corresponds to material torsion and must vanish too, so,
\begin{equation}
\begin{split}
\e_{ik,k} &= 0 \\
\therefore \e^{\mu\alpha}_{\;\;\;,\alpha} &= (1/2)\e^{\alpha\;\;\mu}_{\;\alpha,}
\end{split}
\end{equation}

Since $\e_{\mu\nu}$ satisfies the harmonic gauge condition, we can apply the linearized approximation for the Ricci tensor, 
\begin{equation}
R_{\mu\nu} \approx -\e_{\mu\nu,\alpha}^{\;\;\;\;\;\alpha}
\end{equation}
After substituting into the Einstein Field Equations \eqref{eq:gravity-law}, and taking into account that $R \approx \e^{\alpha\;\;\mu}_{\;\alpha,\;\mu} = 0$, and that in empty space $T_{\mu\nu} = 0$, we arrive at,
\begin{equation}\label{eq:wave}
\begin{split}
\e_{\mu\nu,\alpha}^{\;\;\;\;\;\alpha} & =\e_{ij,kk} - \e_{ij, 00} = 0 \\
\therefore & \e_{ij, 00} = \e_{ij,kk}
\end{split}
\end{equation}
which is a wave equation with solutions that are traveling waves at the speed of light $c$. To see this clearly, let us re-write \Eref{eq:wave} in terms of the coordinate time variable $t$, where $x^0 \equiv ct$, and using the canonical form derivative operators $\partial$, and~$\nabla$,
\begin{equation}\label{eq:wave-long-hand}
\frac{1}{c^2} \frac{\partial^2}{ \partial t^2} \e_{ij} = \nabla^2 \e_{ij}
\end{equation}

The above equation can be related to the Solid Mechanics equation for the propagation of a shear wave in elastic medium with density $\rho$ and shear modulus~$\mu$. In the absence of body forces, the equation of motion is the following,
\begin{equation}
\rho \frac{\partial^2}{\partial t^2} u_i  = \sigma_{ij,j}
\end{equation}
Applying Hooke's Law \eqref{eq:hookes-law} and recognizing that, $\e_{ij} = (u_{i,j} + u_{j,i})/2$, $\mu = Y/[2(\nu+1)]$, and $u_{k,k} = \e_{kk} = 0$, we arrive at,
\begin{equation}\label{eq:s-wave}
\begin{split}
\rho \frac{\partial^2}{\partial t^2} u_i  & = \mu \nabla^2 u_i \\ 
\rho \frac{\partial^2}{\partial t^2} \left(u_{i,j} + u_{j,i}\right) & = \mu \nabla^2 \left( u_{i,j} + u_{j,i}\right) \\
\therefore \rho \frac{\partial^2}{\partial t^2} \e_{ij}  & = \mu \nabla^2 \e_{ij}
\end{split}
\end{equation}
The parallel between Equations \eqref{eq:wave-long-hand} and \eqref{eq:s-wave} confirms that gravitational waves are analogous to shear waves propagating in a solid material and that furthermore the speed of propagation, which is the speed of light $c$, is related to the shear modulus and density of the medium per $c^2 = \mu/\rho$.

Although \Eref{eq:wave} suggests that there are ostensibly ten strain components, $\varepsilon_{\alpha\beta}$, oscillating independently, in reality only two are independent and the rest are coupled to the two. To show this, consider a traveling wave, which corresponds to a gravity wave, propagating along the $x^3$ direction. It is necessary that $\varepsilon_{3\alpha} = \varepsilon_{\alpha 3} = 0$ for the wave to be a shear wave. Furthermore, as shown previously, $\varepsilon_{00} = \varepsilon^\text{3D} = 0$ and $\e_{j0} = \e_{0j}=0$. Finally, we have $\varepsilon^\text{3D} = \varepsilon_{11} + \varepsilon_{22} = 0$, because $\varepsilon_{33} = 0$ already. Therefore, 
\begin{equation}\label{eq:polarizations}
\begin{split}
\varepsilon_{11} &= -\varepsilon_{22} \\
\varepsilon_{12} &= \varepsilon_{21}
\end{split}
\end{equation}
are the only two independent degrees of freedom left, which implies just two types of wave polarizations. The fact that \Eref{eq:polarizations} is in terms of the material strain, which has a definite physical meaning, ensures that the waves must also be physical as opposed to being mere coordinate displacements. This result, derived from a Solid Mechanic's perspective, is consistent with the analogous result from General Relativity about the polarization of gravitational waves~\cite{Misner1973Ch18}. 

\subsection{Elastic Modulus and Density of Free Space}\label{ssec:modulus}

From the result in \Eref{eq:identification}, the fabric's elastic modulus $Y$ could be computed given an estimate for the fabric's thickness $L$. As reasoned in \Sref{ssec:hyperplate},  Planck's length $l_p \equiv \sqrt{\hbar G/c^3}$ is a suitable estimate for $L$, where $\hbar$ is the reduced Planck's constant. Assuming $L \sim l_p$, we can estimate $Y$ to be,
\begin{equation}
\begin{split}
Y \sim \frac{24}{l^2_p \kappa} = 4.4 \times 10^{113} \text{N m}^{-2} \\
\end{split}
\end{equation}
	
The density of the fabric $\rho$ is related to the wave speed and shear modulus, as shown in \Sref{ssec:wave-equation}, and can now be computed,
\begin{equation}
\rho = \frac{\mu}{c^2} = \frac{Y}{4c^2} \sim 1.3 \times 10^{96} \mathrm{kg}\,\mathrm{m}^{-3}
\end{equation}
In accordance with the Cosmic Fabric analogy, the density of the fabric corresponds to the density of free space, which is also known as the zero-point energy density. The computed value for $\rho$ agrees to an order of magnitude with the predictions of Quantum Field Theory ($\sim 10^{96} \mathrm{kg}\,\mathrm{m}^{-3}$) for the energy density of free space~\cite{Rugh2002}. Note that the predictions of Quantum Field Theory are also based on using Planck's length $l_p$ as a length-scale parameter.

\subsection{Generalizing the Cosmic Fabric Model}\label{ssec:generalization}

The Cosmic Fabric analogy to physical space was demonstrated subject to certain simplifying conditions such as small strain (weak gravity), nearly static equilibrium (slow fields) and outside of inclusions (free space). In this subsection, we consider how each of these conditions might be removed or relaxed in a future generalization of the model.

The small strain (weak gravity) condition was imposed so we could use the linearized equations for strain and, analogously, the linearized equations for gravity. To relax this condition, we will need to account for the higher order terms in the equations of strain and also use covariant derivatives instead of conventional differentiation for all field variables.

Imposing the nearly static field condition allowed us to ignore the kinetic energy term in the fabric's Lagrangian. It also let us assume specific bending kinematics that minimize the membrane energy of the fabric and, in \Sref{ssec:membrane-energy}, we showed that such kinematics result in zero membrane energy. Without the condition of nearly static fields, we will need to take into account the kinetic and membrane energies of the fabric and also consider a more complex deformation state. One possible simplification would be to concentrate on the condition without bending (away from static gravitational fields), and derive a closed-form result for the fabric's Lagrangian. This is the condition under which we would study gravitational waves as discussed in \Sref{ssec:wave-equation}. The mathematical complexities resulting in the general non-static fields case will probably require the use of numerical techniques. 

Focusing on deriving the fabric's Lagrangian outside of inclusions was a useful simplification so the $\e_{00,kk}$ term in \Eref{eq:ricci} could be eliminated. In the case of inclusions, this term contributes to an additional Lagrangian term that is analogous to the energy-matter Lagrangian term $\LM$ in the generalized Einstein-Hilbert action $\SEH = \int (R/2\kappa + \LM)\sqrt{|g|}\dif x^4$. The detailed analysis of the fabric's behavior within inclusions will be presented in a subsequent paper.

\section{Summary and Conclusion} \label{sec:summary}

\begin{table*}
	\caption{Comparison between the General Relativity and Solid Mechanics Perspectives.}
	\label{tab:gr-vs-sm}

	\begin{tabular}{p{0.45\linewidth}p{0.45\linewidth}}
		\firsttabline
		General Relativity Perspective & Solid Mechanics Perspective  \\
		\firsttabline
		
		\bodytabline
		
		Physical space &  Mid-hypersurface of a hyperplate called ``cosmic fabric''. \\
		
		\bodytabline
		Spacetime & The world volume of the cosmic fabric's mid-hypersurface  \\
		
		\bodytabline
		Intrinsic curvature of physical space & Intrinsic curvature of the fabric's mid-hypersurface \\		
		
		\bodytabline
		Intrinsic curvature of spacetime & Intrinsic curvature of the fabric's world volume \\		
		
		\bodytabline
		Gravitational potential $\Phi$ & Volumetric strain $\ethree$, such that $\ethree = -\Phi/c^2$ \\		
		
		\bodytabline
		Gravitational waves & Shear waves traveling at the speed of light \\		
		
		\bodytabline
		Matter curves spacetime & Matter induces prescribed strain causing the fabric to bend \\		
		
		\bodytabline
		Action integral in free space, \[ \mathcal{S} = \frac{1}{2\kappa} \int R\sqrt{|g|}\,d^4x \]&  Action integral outside of inclusions, \[ \mathcal{S} = \frac{L^2 Y}{24} \int R\sqrt{|g|}\,d^4x   \]\\

		\bodytabline
		Constants of Nature: \[G,\;\hbar,\;c\] & Elastic constants: \[Y = 6c^7/ 2\pi\hbar G^2,\quad \nu=1 \] \\		
		
		\lastabline
	\end{tabular}
\end{table*}

In this paper, we showed that the behavior of spacetime per Einstein's Field Equations  \eqref{eq:gravity-law} is analogous to that of an appropriately chosen material body, termed the ``cosmic fabric'' that is governed by a simple constitutive relation per Hooke's Law~\eqref{eq:hookes-law}. In \Sref{sec:formulation}, we postulated several basic properties of the fabric and how they correspond to physical space and matter in space. Constitutive properties, such as the Poisson ratio and the elastic modulus dependence on strain, were left unconstrained as model parameters. These were subsequently chosen such that the Lagrangian of the fabric could take the form of the Einstein-Hilbert Lagrangian of General Relativity. After the Cosmic Fabric model was calibrated in this way, in \Sref{sec:discussion}, it was applied to interpret various properties of gravity in terms of the fabric's mechanics and vice versa. To a great extent, the interpretations seemed physically meaningful from both perspectives of General Relativity and Solid Mechanics. \Tref{tab:gr-vs-sm} summarizes the correspondence between concepts from one field to analogous concepts in the other.

The research presented in this paper suggests an equivalence between postulating the field equations of General Relativity and postulating a cosmic fabric having material-like properties as described here. We believe that these are two different approaches for studying the same underlying reality. The Cosmic Fabric model introduces a new paradigm for interpreting cosmological observations based on well-established ideas from Solid Mechanics. In recent decades, Solid Mechanics has made significant advancements in describing the structure of materials at various length and time scales ranging from electrons to large scale engineering structures~\cite{Horstemeyer2000,Horstemeyer2012}. Such advances, in conjunction with the advances in high-performance computing, have made possible the construction of multiscale models that accurately simulate the behavior of metals, ceramics, polymers, and biomaterials. The Cosmic Fabric model should enable the application of these techniques to simulate both the fine and large scale structures of the cosmos, and consequently, to address some of the outstanding problems in Cosmology, such as those pertaining to the density of free space, dark energy, and dark matter. 

In this paper we focused on weak gravitational fields that are exterior to gravitating bodies. Within this scope, we further focused on nearly static fields, and separately, on gravitational waves in the context of nearly flat spacetime. In our continuing research, we have also applied the model presented herein to the exterior of gravitating bodies as well as to moving bodies, and we have found that the same postulates continue to yield results that are consistent with the Theory of Relativity, and we hope to publish them in the near future. 

Our research in developing and applying the Cosmic Fabric model is still ongoing, and the results we have shared here are subject to some simplifying constraints. In the future, we hope to relax these assumptions if not eliminate them completely. Nevertheless, even at the current stage of the work these results appear promising and useful.

\section*{Acknowledgments}

MFH would like to acknowledge the Center for Advanced Vehicular Systems (CAVS) at Mississippi State University for support of this work. The authors would also like to thank Russ Humphreys and Anzhong Wang for their insight with respect to Cosmology, and Shantia Yarahmadian for his feedback with respect to Mathematics. TGT would also like to acknowledge Tamas Morvai for the insightful discussions on the nature of time.

\bibliographystyle{ws-ijmpd}
\bibliography{mechanics-of-spacetime}

\nomenclature[01]{$Y$}{Young's modulus of elasticity.}
\nomenclature[02]{$\nu$}{Poisson's ratio (when not used as an index).}
\nomenclature[04]{L}{Reference thickness of the cosmic fabric.}
\nomenclature[05]{$\hbar$ }{The reduced Planck constant, $\hbar=1.054571800\times 10^{-34}  \mathrm{m} ^2 \mathrm{kg}\, \mathrm{s}^{-1}$ .}
\nomenclature[06]{$G$}{The gravitational constant,  $G=6.67408 \times 10^{-11} \mathrm{m}^3\,\mathrm{kg}^{-1}\mathrm{s}^{-2}$ .}
\nomenclature[06.1]{$c$}{Speed of lgiht, $c = 299,792,458\, \text{m s}^{-1}$}
\nomenclature[07]{$\kappa$}{The Einstein constant, $\kappa \equiv 8\pi G / c^4 $.}
\nomenclature[09]{$\nabla^{2}$}{The Laplace operator for spatial coordinates, $\nabla^{2}\equiv\partial_{11}+\partial_{22}+\partial_{33}$. }
\nomenclature[11]{$x^{a},\:x_{a}$}{Upper indexes indicate contravariant components; lower indexes indicate covariant components. }
\nomenclature[12]{$x^{a}y_{a}$}{Contraction on repeating indexes: $x^{a}y_{a}\equiv\sum_{a}x^{a}y_{a}$. Indexes must appear in opposite top versus bottom position. }
\nomenclature[13]{$\delta_{b}^{a}$}{Kronecker delta: $\delta_{a}^{b}=\delta_{ab}=\delta^{ab}=\{1,\:a=b;\quad 0,\:a\text{\ensuremath{\neq}}b \}$. }
\nomenclature[14]{$g_{\mu\nu},\:g$}{The spacetime metric and its determinant, $g \equiv \det g_{\mu\nu}$.}
\nomenclature[16.1]{$\eta_{\mu\nu}$}{The Minkowski metric for flat spacetime, $\eta_{00}=-1,\:\eta_{0i}=\eta_{i0}=0,\:\eta_{ij}=\delta_{ij}$.}
\nomenclature[17]{$\varepsilon_{ij},\,\varepsilon^{3D}$}{Small strain tensor and its trace, $\varepsilon^{3D}\equiv g^{ij}\varepsilon_{ij}$ in three-dimensions.}
\nomenclature[18]{$\varepsilon_{\mu\nu},\,\varepsilon$}{Small strain tensor and its trace, $\varepsilon \equiv g^{\mu\nu}\varepsilon_{\mu\nu}$ in four-dimensions.}
\nomenclature[19]{$\sigma_{ij},:\sigma_{\mu\nu}$}{Stress tensors in space and spacetime, respectively.}
\nomenclature[19.1]{$\sB_{ij},\:\sB_{\mu\nu}$}{Stress tensors due to bending in space and spacetime, respectively.}
\nomenclature[19.2]{$\sM_{\mu\nu}$}{Stress tensor due to matter-energy inclusions.}
\nomenclature[20]{$R_{\mu\nu},\,R$}{The four-dimensional Ricci curvature tensor and scalar, $R\equiv g^{\mu\nu}R_{\mu\nu}$.}
\nomenclature[21]{$R_{ij}^{3D},\,R^{3D}$}{The three-dimensional Ricci curvature tensor and scalar, $R^{3D}\equiv g^{ij}R_{ij}^{3D}$.}
\nomenclature[22]{$T_{\mu\nu},\,T$}{The stress-energy tensor of General Relativity and its trace, $T\equiv g^{\mu\nu}T_{\mu\nu}$.}
\nomenclature[23]{$\textrm{diag}[a_{00},\,a_{11},\,a_{22},\,a_{33}]$}{ Denotes a $4\times4$ matrix with diagonal elements $a_{00},\dots,a_{33}$ and vanishing off-diagonal elements.}


\end{document}